**Virus - Encoded Ribonucleotide Reductases**

Brief Review

Dr. Claus Bornemann
Eurofins Umwelt West GmbH
Vorgebirgsstr. 20
50389 Wesseling
Germany
clausbornemann@eurofins.de

**Ribonucleotide Reductases**

Ribonucleotide reductases [EC 1.17.4.] are a group of enzymes which critically contribute to the biosynthesis of deoxyribonucleotides (dNTPs), the monomeric precursors of DNA polymerization. By a radical reaction mechanism, they catalyze the reduction of the ribonucleotides, ADP, GDP, CDP, and UDP, mostly at the diphosphate level, to their deoxy conterparts dADP, dGDP, dCDP, and dUDP, respectively. Together with other enzymes, i.e. nucleoside and nucleotide kinases and phosphatases, deaminases, methylases and hydroxymethylases they are part of the biochemical pathway ultimately leading to dATP, dGTP, dCTP, and dTTP, the final substrates of DNA polymerases [31].

Of critical importance is not the amount of dNTPs, but are the relative concentrations of the four at the replication forks. Elongating and proofreading activities of DNA polymerases are affected by alterations of these concentrations, and biased dNTP concentrations are mutagenic. The enzymes in the deoxynucleotide biosynthetic pathway, especially ribonucleotide reductase which reduces all four ribonucleotide substrates, underlie extensive allosteric regulations with the effect to bring about a balanced and species-specific supply of dNTPs [18, 24]. Ribonucleotide reductases have a medium-complex subunit organization with an elaborate array of active and allosteric nucleotide binding sites. Most well-known is the $\alpha_2\beta_2$ iron-dependent tyrosyl radical specimen of aerobic E. coli and eukaryotes [31]. Mutants of E.coli ribonucleotide reductase have been engineered which produce different deoxynucleotide ratios and act as mutator [1]. They induce base replacements in the direction of the nucleotides which they provide in excess.

There are different classes of ribonucleotide reductases in which the functionally important, protein side chain based radical is obtained by an upstream reaction sequence involving diverse metal cofactors [27, 35]. Radical acquisition can be either oxygen-dependent, oxygen-independent, or oxygen-sensitive, i.e. strictly anaerobic [39]. Reduction equivalents are provided by thioredoxins and glutaredoxins [19]. Deoxynucleotide synthesizing enzymes are moderately cell-cycle regulated. This can most significantly be observed in unicellular green algae grown in a 24 h light-dark-regime. In the dark, cells enter $G_0$ phase, and ribonucleotide reductase activity vanishes [4, 14].

**Viral Ribonucleotide Reductase Genes and Enzymes**

Genes for ribonucleotide reductases are found in most viruses with genome sizes of 100 kbp and more. The first and best studied examples are in E.coli bacteriophage T4 [29, 39], Herpes viruses [12, 30], and the poxvirus, Vaccinia virus [34, 36]. In these cases, virus-coded ribonucleotide reductases have been expressed, separated from their host cell counterparts, and their enzyme properties have been characterized [2, 3, 17, 18, 22]. The allosteric properties, especially the relative yields of the four products dADP, dGDP, dCDP, and dUDP, have been found to differ from those of the host enzymes in T-phage and Herpes viruses, but not in Vaccinia virus.

Significantly, only a few viruses like phage T4 encode a complete deoxynucleotide biosynthetic pathway [29]. It even possesses, like its host, an alternative anaerobic ribonucleotide reductase [39]. Most viruses do not play with a full deck [37], and some, e.g. the herpesvirus Human Cytomegalovirus, encode only one subunit of ribonucleotide reductase, which on its own is probably non-functional [8, 9].

Genes with sequence homology to ribonucleotide reductase continue to be found in baculoviruses [21, 38], African Swine Fever Virus [5], Chlorella Virus [25], Mimivirus [33], and, recently, in over 100 more bacteriophages [13], with cases again of lone subunits and with degenerating genes among them.

**Functions of Viral Ribonucleotide Reductases**

With accumulating DNA sequence data suggesting that viruses are rather gene shuttles than functional units and as it becomes unclear whether they are under selective pressure all the time [16] one is reluctant to ponder on possible functions of virus-specified genes. Viruses surely contain a core set of genes which are functional and indispensable for their maintenance and propagation, e.g. those for shell proteins. Small viruses like Simian virus 40 and adenoviruses further possess multifunctional gene products like their large T antigen or E1A which induce and recruit host cell functions needed for replication [11]. Large DNA viruses on top of this basic inventory encode functional homologues of cellular proteins. In Herpes and pox viruses, some of these are probably useful for evasion from the host immunosystem [6, 32].

Genes encoding enzymes of deoxyribonucleotide metabolism can in principle only be "functional", in the sense of conferring independence from the host, if the pathway is complete. In most large viruses this is not the case. Especially thymidylate synthase which is indispensable for the formation of dTMP from dUMP is rarely encoded by viruses [37]. Viral ribonucleotide reductase genes are often obviously functionless, i.e. lone or degenerate [13].

When discovered, viral ribonucleotide reductases were regarded as possible therapeutic targets. Considerable effort went into elucidation of the significance of Herpesvirus ribonucleotide reductases [2, 10, 15] and their use as an Achilles heel in chemotherapy. All these attempts gave no useful results [23], obviously because virus-specified ribonucleotide reduction is not limiting in the infective cycle. Deoxynucleotides are readily provided by the host cell dNTP-synthesizing machinery which is induced during infection ("S Phase-like environment" [7]).

If we lavishly calculate dNTPs needed for replication of a virus with a large burst size and 90% waste DNA (i.e. not properly resolved and not packaged into virions), we arrive at 200 kBp x 1000 plaque forming units/cell x 10 which gives $2 \times 10^9$ base pairs. This number corresponds to an average eukaryotic genome size. So there is no demand for higher amounts of deoxynucleotides upon virus infection.

In conclusion virus – encoded ribonucleotide reductases rarely function as the sole and indispensable source of the deoxynucleotides needed for DNA replication. They merely "interfere with nucleic acid metabolism" [37].

**Possible Effects of Viral Ribonucleotide Reductases**

Nonetheless, virus – encoded ribonucleotide reductases are expressed during the infective cycle along with the host enzymes and will contribute to and alter dNTP pools according to their specific and sometimes different regulatory properties. The guanine + cytosine vs. adenine + thymine molar ratios of Herpesvirus genomes encompass the whole viable range. Honess [20] suggested that this be due to a non selective mutation mechanism probably residing in their nucleotide metabolism. This could possibly be the allosteric properties of ribonucleotide reductase. For Herpes simplex virus 1 the low Michaelis constants of its ribonucleotide reductase for GDP and CDP [2] are in accord with the high (70%) guanine + cytosine content of its genome, and for other viruses this should be testable. For example two baculoviruses, Cydia pomonella granulovirus (CpGV), and Cryptophlebia leucotreta granulovirus (CrleGV), are closely related and highly homologous at the protein level but differ in the adenine + thymidine content of their DNA (55 vs 68%) [28]. This difference is mainly reflected in the wobble position of codons. CpGV encodes a ribonucleotide reductase, and CrleGV does not.

When the base composition of a viral genome is driven to an extreme where most further base replacements in the biased direction would result in non-functional proteins and thus less viable offspring, virus coded ribonucleotide reductases with mutator properties may exert a selective advantage. Further base replacements in the biased direction would be fatal, and base replacements against the biased direction are disfavoured, because these nucleotides are already at shortage. Thus, in a biotope with a given number and array of infectable host cells, a smaller number of still possible mutations can be probed more effectively for further development. There would be less competition with an otherwise vast number of neutral mutations.

Thus, forcing a GC/AT bias as extreme as possible is like erecting a wall. It would be like building a stable borderline along which the genome can evolve. The mutating force would be the altered deoxynucleotide concentrations brought about by a differently regulated ribonucleotide reductase and the selective force would be the maintenance or improvement of biological function of the proteins encoded by the GC/AT biased viral genome. Completely detached from their commonly assumed function, viral riboncleotide reductases would then initially be non-selective Honessian mutators. Yet, as they drive them to their base composition extreme, they would gradually make their viruses quicker, more aggressive evolvers.

**Acknowledgements**